\documentclass[11pt]{article}
\usepackage{moriond,epsfig}

\def\Journal#1#2#3#4{{#1} {\bf #2}, #3 (#4)}

\def\PRL{\em Phys. Rev. Lett.}
\def\PRD{{\em Phys. Rev.} D}

\def\APJ{\em Astrophys. J.}
\def\AP{\em Astroparticle Phys.}

\def\be{\begin{equation}}
\def\ee{\end{equation}}
\def\bea{\begin{eqnarray}}
\def\eea{\end{eqnarray}}

\begin{document}
\vspace*{4cm}
\title{SUMMARY OF THE XXXIX RENCONTRES DE MORIOND}

\author{ M. ROOS}

\address{Department of Physical Sciences, PL 64,
FIN-00014 University of Helsinki,\\ Helsinki, Finland }

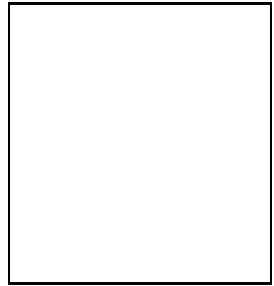
\maketitle\abstracts{This conference covered dark matter particle properties and searches, lensing by dark matter, and dark matter distributions in galaxies and clusters; dark energy phenomenology and theoretical models; surveys of structure and galaxy formation, large scale simulations and neutrino cosmology; astronomy and star formation; cosmic rays; cosmic microwave background; inflation in the primordial universe, cosmic strings, brane cosmology, and the fine-structure constant.}

\section{Introduction}
Although I had only about half a minute per speaker I am consoled
by the fact that every speaker had 20 minutes -- the review
speakers 40 minutes -- so my additional half a minute or not does
not add or subtract much. In this situation some summary speakers
would abandon their task and speak only about their own research
-- you have to be very famous to do that, the Nobel prize is a
common excuse -- or then resort to telling jokes, but then only
good ones. Three categories of talks I find particularly
difficult to comment meaningfully are review talks which already are
summaries, large simulations, and the intricacies of ingenious
detector designs.

\section{Dark Matter}
Whatever the DM (Dark Matter) is, it is endowed with mass that interacts gravitationally with conventional mass, and thereby it influences the dynamics of massive structures and the time scales of their formation. There are independent evidences for the existence of DM on all scales, from dwarf galaxies to superclusters to the global properties of the known Universe  (see {\em e.g.} Roos~\cite{ro}). The fraction of matter in the Universe is expressed by the mass density parameter $\Omega_m$, which contains the dark matter fraction, $\Omega_{DM}$, the relatively small contribution of baryonic matter, $\Omega_b$, and the minute contribution of neutrinos, $\Omega_{\nu}$.

\subsection{Particle Properties and Searches}\label{subsec:part}
{\em P. Fayet} opened the conference with the proposal of a light DM particle with mass and interactions constrained by relic abundance and structure formation rather than by theory. A viable possibility could be a new light gauge boson with a new annihilation mechanism and with vector coupling to ordinary matter. The annihilation cross-section would be stronger than weak interactions at low energies and energy-dependent. A signal would be low-energy $\gamma$-rays from the GC (Galactic Center), maybe the positrons causing the 511 keV line. This is an example of an idea which is 'cheaper' than SUSY neutralinos, $\chi$, because SUSY implies the existence of some 50 new sparticles which nobody has seen.

WIMP detection -- actually non-detection -- is a very active field with many currently running detectors and future ones. The most popular WIMP candidate is the $\chi$  which is expected to scatter elastically against nucleons, or to annihilate into detectable photons. {\em G. Chardin} summarized the signatures of SUSY models, WIMPs left-over from the thermal equilibrium and the experimental techniques to find them. 'Still desperately looking for SUSY' he said, but I think there are actually two desperations: looking for DM particles and looking for SUSY. My prediction is that DM will be understood sooner than SUSY is proved. Chardin further discussed the experiments DAMA, Edelweiss-I, CDMS-I, CRESST-II, and ZEPLIN-I and concluded that we are at last becoming sensitive to SUSY models, but next generation experiments with a factor $\approx 50$ improvement in sensitivity will allow to test the bulk of SUSY parameter space.  

{\em P. Brink} reported the first results from the cryogenic CDMS-II in the Soudan underground lab. This sets stronger exclusion limits SUSY neutralinos than any previous experiment. In particular it is incompatible with DAMA's independent annual-modulation data at $>99.8 \%$ confidence. Similarly, {\em V. Sanglard} presented a preliminary Edelweiss elastic WIMP-nucleon scattering cross-section limit which also excludes the DAMA $3\sigma$ region completely, confirming the Edelweiss 2002 exclusion limit. A problem is, however, to understand what DAMA means by $3\sigma$: the comparison of confidence regions in different experiments is not necessarily consistent. {\em V. Kudryavtsev} showed a preliminary limit on the WIMP-nucleon scattering cross-section from the ZEPLIN I liquid Xe TPC-detector at the Boulby Mine. This limit also excludes the DAMA $3\sigma$ region completely. ZEPLIN II and III are being commissioned, the DRIFT I detector is installed and is running. {\em E. Moulin} described the MACHe3 detector which will be sensitive to low energy conversion of electrons from non-baryonic DM with axial interactions.

{\em S. Colafrancesco} gave a splendid cultural show of uses and misuses of the term Dark Matter outside physics. He reviewed the detection of $\chi$ annihilation in galaxy clusters by secondary high-energy electrons, by the SZ (Sunyaev-Zel'dovich) effect of CMB scattering against electrons causing $\gamma$ pressure, and in the radio-, X-ray, and $\gamma$-ray spectra. He concluded that nothing can be said at present about a $\chi$ with mass exceeding 35 GeV.

{\em L. Pieri} studied the possibility to distinguish $\chi$ annihilation into $W^+\ W^-$  and into photons in the local group galaxies M31, M87, LMC, and SMC above the Galactic background, with future detectors in space, in particular in the presence of sub-halo clumping and black holes. {\em F.D. Steffen} proposed that stable axinos, left-over from post-inflationary thermal scatterings at $T_R<10^6$ GeV, might be the dominant DM. Another DM idea was proposed by {\em F. Piazza}: DM might interact with baryons with a Yukawa-type force cancelling gravitation at kpc distances. This would help to explain the DM profiles in Low Surface Brightness galaxies. {\em L. Derome} discussed the upcoming (2007) magnetic spectrometer AMS-02 that would be sensitive to 0.5 -- 3 GeV DM particles decaying into $\bar{p}$ and $e^+$, and to the presence of $\overline{\rm He}$. Thus it could detect possible inhomogeneous baryosynthesis. 

{\em A. Lionetto} described GLAST which will be looking for annihilation photons ($E_{\gamma}< 300$ GeV) in the GC, and noted that the GC excess seen by EGRET could be explained as an effect of SUSY  and a moderate halo profile. GLAST will allow to understand if the GC-source is due to standard processes or to MSSM or mSUGRA. However, {\em F. Ferrer} argued that $\chi$ annihilation into photons in the GC is an unpromising venture since the DM distribution is not cuspy.  A better place to look would be Sagittarius or Draco which have a high $M/L$. {\em W. Wittek} described the now installing MAGIC imaging Cerenkov telescope which will be sensitive to $E_{\gamma}=30-200$ GeV photons. He also noted that de detection of a DM signal from the GC is unlikely, whereas Draco and Sagittarius remained candidates.

In spite of this broad activity of searches, no candidate has yet been identified.

\subsection{Lensing}
Let me first clarify lensing terminology a bit (see {\em e.g.} Roos~\cite{ro}). The {\em lens} or the {\em lensing object} may be a star, a dark object, a galaxy, or any mass inhomogeneity that curves space locally. Behind it is the {\em source} or the {\em lensed object}: a star, a QSO, or a galaxy, an image of which we observe through the curved space of the lens. If  the lens and the source are on a nearly straight line from us, one has the case of {\em strong lensing} which produces several (an odd number of) distorted pictures of the source, giant {\em arcs} and {\em caustics}. In the case of an exactly straight-line geometry, the caustics form a symmetric {\em Einstein cross} which collapses to an {\em Einstein ring} when the radius of the lens is small. {\em Microlensing} corresponds to the case when the resolution is smaller than the radius of the Einstein ring. The image is then seen magnified during the time the lens passes in front of the source. {\em Screen-lensing} denotes the case when the source is distant, but the lens is within our halo. {\em Self-lensing} denotes the case when both the source and the lens are in the same distant galaxy. If the lens and the source are not on a straight line joining us, no caustics are formed, only elliptical deformations of the image of the source. This is called {\em weak lensing},  an important tool to study the large-scale distribution of gravitating matter. Ellipticity and shear can enhance the magnification.

{\em O. Perdereau} reviewed present microlensing results: geometries, amplification, distortions, targets (the GC, the spiral arms, LMC, SMC, and further away). There are now five surveys which have found $\approx 20$ events in the Galactic halo towards the LMC/SMC. The LMC and SMC exhibit some differences which indicate that microlensing may be a means to test galaxy structures. There are strong constraints on halo Machos of mass below $0.1$ M$_{\odot}$. This is a very active area: OGLE-III and SuperMacho are starting, several groups are looking at the M31, and there are several projects in space and on ground. 

{\em A. Rest} reported the first results from SuperMacho, a next-generation microlensing survey towards the LMC with the goal of determining the location of lensing populations: discriminating self-lensing from screen-lensing by the differential event rate.  So far 50 candidate microlensing candidates have been found. Microlensing events have to be distinguished from other confusing variable stars, notably SNe (supernovae) behind the LMC, of which 70 have been found in 2003. In the near future a better event classification can be expected, with more spectroscopic/photometric follow-up.

Looking at more distant galaxies, individual microlensing stars cannot be isolated, only microlensing pixels. {\em S. Paulin-Henriksson} reported that the  POINT-AGAPE collaboration had found seven microlensing pixels towards the M31, all bright and of short duration. This excludes that the halo should contain more than 25\% dark objects of mass $0.1-0.6$ M$_{\odot}$. {\em A. Zakharov} reported on calculations of the optical depth of deflectors in the bulge and halo of host galaxies, and of cosmologically distributed deflectors, pointing out that stellar mass DM in QSO bulges and halos could be observed by X-ray microlensing much more readily than in the optical or UV. {\em A. Amblard} had simulated weak lensing DM fields and kinetic SZ fields (kSZ) which create a non-Gaussian bias in the CMB anisotropy maps and in the power spectrum. Studying the effect of kSZ fields, foreground effects and non-Gaussianity on random realizations of CMB temperature anisotropies, he showed that the kSZ bias could be removed with suitable masks.

\subsection{Distributions in Galaxies and Clusters}

Precise knowledge of the DM profile in galaxies and clusters would give valuable information about the interactions of DM.  {\em W. Wittek},
already mentioned in section ~\ref{subsec:part}, discussed extensively the form of DM profiles, in particular in the GC and in Draco and Sagittarius. {\em D. Kazakov} reported on the reconstruction of the DM profile in the Galaxy and of its rotation, from EGRET observations of excess diffuse $\gamma$-rays above 1 GeV. The same spectrum in all regions of the sky map suggests a common origin, such as neutralino annihilation at $E_{\chi}=100-200$ GeV. This is of course model-dependent because it assumes that the photons indeed come from DM.  The DM halo is then found to be ellipsoidal, less cuspy than the NFW (Navarro-Frenk-White) model and the Moore model. Peaks in the $\gamma$-ray angular distributions can be described by DM clumps similar to visible matter clustering, producing the spiral arms and indicating DM rotation. Photons, positrons, and antiprotons determine a single halo profile with the same MSSM parameters; this is stated to be an interesting hint that DM may be the SUSY partner of CMB.

Turning now to galaxy clusters, {\em M. Lombardi} reported on a weak lensing observation with the HST/ACS instrument, at $z = 1.24$. The mass profile of this cluster agrees with previous observations by CHANDRA and XMM-Newton. 

From a large study of the collisional effects in $10^5$ time steps of 25M particles inside the virial radii of six simulated clusters, {\em J. Diemand} reported that the CDM halos could not be fitted with either the NFW 2-parameter profile nor Moore's profile. Inside  0.3\%  of the virial radius the baryon density spoils all CDM density fits. Instead, a 3-parameter NFW-like profile fitted well for radii down to 0.3\%  of the respective virial radius. This is not surprising: the NFW model is essentially phenomenological, so with increasing statistics it will need more parameters for a good description. Moreover, matter is clumpy, Diemand identified $<5000$ subhalos. The simulated clusters still show disagreemnt with observed clusters; could hydrodynamics explain this or is it a real problem for CDM ? 
In another study of the statistics of substructure in a large N-body simulation, {\em J. Weller} developed an algorithm allowing the identification and dynamical analysis of structures in DM clusters. He concluded that the mass fraction in substructures has a large variance.

A combination of strong and weak lensing probes the matter density in an interesting way. The visible part of a cluster can be studied with strong lensing and photometry, whereas weak lensing gives independent information at radial distances where photometry is difficult. Probabilistic analysis of complex cluster lenses is complicated by the non-circular shapes of the lenses, as shown by {\em P. Marshall}. The majority of clusters is likely to have substructure that needs to be accounted for when measuring halo properties. 
{\em M. Bradac} also combined weak and strong lensing, distortion and redshift information of background galaxies for cluster mass reconstruction. The convergence parameter $\kappa$ in the Jacobian matrix between the source-plane coordinates and the observer-plane coordinates (see {\em e.g.} Roos~\cite{ro}) can be determined from weak-lensing ellipticites if one knows the redshift; if not, $\kappa$ is affected by a mass-sheet degeneracy which can only be broken for strong-lensing clusters capable of multiple imaging. Thus strong lensing and weak lensing complete each other.

\section{Dark Energy}
Historically the first evidence for repulsive DE (Dark Energy) came from the observations of distant SNe Ia. The contribution of DE to the total energy of the Universe is expressed in the concordance model which has a cosmological constant $\Lambda$, by $\Omega_{\Lambda}$. In a flat universe, $\Omega_m+\Omega_{\Lambda}=1$.

\subsection{Phenomenology}\label{subsec:phen}
In the art of combining different data to constrain sets of parameters, the normally used technique to find best values is the maximum likelihood statistic, a classical frequentist concept. Bayesian statistics, in contrast, considers degrees of belief, converting a prior belief into a posterior belief. The Bayesian and frequentist statistics lead to the same result if the prior for a parameter has a typical likelihood form, say a Gaussian probability density function. A flat prior is dangerous, because flatness is not equivalent to absence of knowledge. A flat distribution for a parameter $P$ is not flat in $P^2$ or in $\log P$. Since data are normally entered as constraints of $e.g.$ Gaussian form, they should not be called priors, but constraints.

{\em M. Giavalisco} reviewed some of the large body of new results from the GOODS collaboration, notably their discovery with the HST/ACS detector of 16 high-$z$ SNe Ia, ranging out to $z=1.8$ (cf. A. Riess {\em et al}~\cite{ri}). The SNe Ia can be distinguished from the SNe II by color and time distribution. The $z$-distribution of the SNe Ia appears to show evidence for an epoch of deceleration which turned into the present accelerated epoch at $z\approx 0.3-0.6$, 5 Gyr ago. Einstein's model with a cosmological constant really looks better than ever. From 80 SNe Ia at $z=0.3-0.9$ and 8 at $z>1.0$ the total mass density parameter is $\Omega_m=0.27\pm 0.04$.

The supernova hunt is thus on. {\em J. Guy} described the first results from the Canada-France-Hawaii Telescope Legacy weak lensing Survey (CFHTLS) which centered on finding SNe and measuring the SN rate as a function of redshift. They aim at 700 SNe Ia in 202 nights of observation. So far the collaboration has identified 38 SNe Ia, 10 further events still needing identification by photometry. Currently they are working on improving detection pipeline, automated off-line photometry, light-curve fitting, etc. {\em G. Sainton} reported on the SNLS collaboration's work on distant SN Ia spectroscopy with the VLT.  They measure the redshift, identify type, age, and environment. The present status after 6 months of observations is 27 SNe Ia identified in the range $0.17<z<0.95$, one SN Ic, some peculiar ones, and more than 50 candidates spectral-analyzed. (Further talks on GOODS and CFHTLS will be summarized in section ~\ref{subsec:survey}.)

One can now start to include also the time-dependence of $w_{\phi}$, the EOS (Equation Of State) of a quintessence field $\phi$, as a parameter in a linear expansion, $w_{\phi}=w_0 + w'_0 z$.  The combination of supernova data, CMB, and LSS (Large Scale Structure) data from the SDSS shows that the information on $w'_0$ is still weak, a large region of $w'_0<0$ values is allowed. {\em A. Melchiorri} continued the discussion of these data, pointing out that the best fit value of a time-independent $w_{\phi}$ is in the 'unphysical' Big Rip region $w_{\phi}<-1$, also called Super-Acceleration or Phantom Energy, where the Weak Energy Condition $p+\rho\geq 0$ is violated. Stated as a limit in the physical region, $w_{\phi} <-0.8$ at 95\% C.L. (not conditional on $w_{\phi}> -1$). But a fit to the form $w_{\phi}=w_0 + w'_0 z$ does not have much likelihood in the parametric regions leading to a Big Rip (cf. Wang and Tegmark~\cite{te}), and the need for a parameter $w'_0$ is not convincing.

{\em S. Bridle} pointed out that even if $w_{\phi}$ is constrained on the average to be physical, it does not exclude temporary excursions into the unphysical region. She also proposed a phenomenological and statistical method to measure $w_{\phi}$ and test whether $w_{\phi}<-1$. The EOS of a cosmological constant, $\Lambda$, differs from that of a quintessence field in being always constant, $w_{\Lambda}=-1$, but the value of $\Lambda$ could vary with time. Such models have been studied in the literature. She concluded that if the Universe super-accelerated enough, Planck should be able to determine if $w_{\phi}> -1$ at some $z$ without making assumptions about the form of $w(z)$ or about $\Omega_m$. {\em P. Ruiz-Lapuente} showed that current SNe Ia spectra show no evolution in $\Lambda$ in the redshift range $z=0.1-0.3$ compared to the present. 

{\em D. Parkinson} pointed out that to distinguish between $\Lambda CDM$ and quintessence one would need better information on the amplitude of LSS fluctuations, $\sigma_8$. From the present data he derived the limit $w_{\phi} <-0.82$, finding no strong change in $w$ for $z<1$ and no improvement in going from $\Lambda CDM$ to QCDM. The reason for this lack of information was neatly spelled out by {\em C. Horellou}: for a constant $w_{\phi}$ in the range $-1\leq w_{\phi}\leq -1/3$, the larger $w_{\phi}$ is, the earlier do clusters form and the more concentrated do they become. But $\sigma_8$ decreases with increasing $w_{\phi}$, and thus the effect of concentration is counteracted. She computed the abundance of clusters for different choices of $w_{\phi}$ and $\sigma_8$, concluding that it is difficult to obtain a value for $w_{\phi}$ from clusters.

\subsection{Theoretical models}
{\em A. Melchiorri} pointed out that the new constraints on DE posed by the GOODS high-$z$ SNe Ia will require modifications either in the field equations or in the matter sector. The big questions are why $\Lambda$ is so small and why it conspires to make DE of the same magnitude as DM  right now, the 'cosmic coincidence problem'. 

Quintessence models with exponential potentials can be treated  general-relativistically exactly, they are tracking and are insensitive to initial conditions when $w$ is sufficiently different from $-1$. The $\Lambda CDM$ model with $w=-1$ requires exact fine-tuning and does not track~\cite{blu}. {\em S. Capozziello} showed that this leads to a $w_{\phi}$ which is maximally variable now; is this another cosmic coincidence ? Such a model is, however, today indistinguishable from the $\Lambda CDM$ model. As a warning to anybody playing with phenomenological potentials he pointed out (as previously noted~\cite{bl}) that the quantity $\Gamma$ in the tracking condition of Steinhardt {\it et al}.~\cite{swz} is wrong. In fact, the condition $\Gamma>1$ seems to be unnecessary. The study of tracker behaviour is subtle and depends strongly on the representation chosen.

Models which relate DM and DE have attracted attention because they may appear to solve the coincidence problem or to alleviate it, perhaps at the expense of some other kind of fine-tuning. {\em A.V. Maccio} demonstrated that a scalar DE field with Ratra-Peebles or SUGRA potentials changes giant lensing arcs number density. The Ratra-Peebles potential is ruled out by recent observations, whereas SUGRA makes correct predictions at all redshifts. {\em G. Huey} presented a model in which quintessence (DE) coupling to a variable mass DM causes the DE to decay rapidly during radiation domination. During matter domination it decays more slowly than the matter density, and so it becomes dominant at some epoch. 

Following the end of inflation, an increased DE density increases the expansion rate $H$, and this causes WIMPs to freeze-out at a higher temperature. On the other hand, the slower the DE potential decreases, the longer is the kination time (cf. also Dimopoulos and Valle~{\cite{di}}). {\em S. Profumo} discussed many scenarios that only depended on two parameters, $H_{end}$ at the end of inflation and $\lambda$, the exponential slope of the DE potential. Since the neutralino is some composite of binos, winos, and higgsinos, one has to keep track of these components separately and the timing. The heavier the freezing-out particle, the larger the DE effects at decoupling. Winos and higgsinos annihilate faster, freeze-out earlier and underproduce DM. Thus the neutralinos are mainly binos. A large annihilation rate leads to a large DM detection rate, but it also leads to a lower relic density.

Turning now to more exotic fields, {\em M. Pietroni} proposed a scalar-tensor theory of gravity with an ultralight quintessence field. The bounds from BBN, CMB and solar system tests of gravity could change the relic DM density substantially in this model, so that the expansion could have been be speeded up by a factor up to $10^5$ at early times. {\em I. Neupane} proposed that DE could be a gravitational scalar potential arising from slowly varying sizes of extra dimensions. This is string or M-theory compactification on hyperbolic spaces. If our Universe is exactly spatially flat, the present accelerating phase is transient and unique, otherwise it could be eternal, or there could be many transient phases of acceleration. 

{\em M. Morikawa} considered DE in the form of a BEC (Bose-Einstein Condensate) of a boson field with attractive interaction leading to negative pressure. A spatially uniform mode of the BEC would cause the global accelerated expansion, whereas a local BEC mode would produce boson stars, black holes, and act as DM. After many cycles of local BEC implosions and sedi\-mentation, boson star and supernova explosions, the Universe would gradually be filled with stars, galaxies, black holes and expanding ordinary hot gas, so that finally the ratio of DM to DE would self-regularize to be of order one. 

{\em S. R\"{a}s\"{a}nen} pointed out that the energy densities on scales $<10$ Mpc are not homogeneous so that our knowledge of the pressure $p$ and the expansion rate $H$ are determined by averaging. But Einstein's equations hold for $p, H$ and $g_{\alpha\beta}$, not for $\langle p\rangle, \langle H \rangle, \langle g_{\alpha\beta}\rangle$. Typically $\langle g_{\alpha\beta}\rangle$ contains terms with $w$ values between $-4/3$ and $-1/3$ which modify Hubble's law and constitute a back-reaction of linear perturbations. This could be the explanation for the observed accelerated expansion.

\section{Structure and Galaxy Formation}

\subsection{Surveys}\label{subsec:survey}
In section ~\ref{subsec:phen} two talks on the surveys GOODS and CFHTLS were already summarized. In a review talk {\em M. Giavalisco} described the GOO Deep Survey which aims to establish deep reference fields with public data sets from X-ray through radio wavelengths for the study of galaxy and AGN evolution of the broadest accessible range of redshift and cosmic time. It unites the data from NASA's space observatories HST, Chandra, SIRTF, and ESA's XMM-Newton as well as from the great ground-based observatories. {\em H. McCracken} reviewed CFHTLS work with XMM-Newton data and Chandra data, surveying the luminosity and temperature of 1M high-redshifts corresponding to some 200 hours of integration time. So far 20000 redshifts have been obtained.

In another review talk {\em Ch. Marinoni} summarized the public VIRMOS-VLT redshift survey that now comprises a multi-color catalog of 3M objects. They measure a strong evolution of the DM biasing function in a flux-limited sample of 4700 galaxies within $0.7<z<1.5$. They reconstruct the galaxy density fields, detect non-linearity, and find that the color-morphology relation was already in place at $z\approx 1.5$. {\em O. Ilbert} described a preliminary sample of 4500 spectroscopic redshifts of galaxies from the VVDS collaboration. These constrain the LF (Luminosity Function) shape and the LF evolution according to the galaxy type and environment up to $z=1.5$. They see LF brightening with increasing $z$ and steepening of slope, mainly in the bluest type galaxies.

The 2dFGRS survey was covered by two talks. {\em P. Norberg} discussed hierarchical scaling in galaxy distributions, showing that volume-averaged $p$-point correlation functions can be written in terms of volume-averaged 2-point correlation functions (variances) if the initial fluctuations are Gaussian. The 2dFGRS exhibits this feature up to $p=6$. But rare and extreme superstructures can strongly influence hierarchical amplitudes. {\em Th. Contini} surveyed the physical properties of galaxies in the range $1.5<z<2.5$. {\em S. Vauclair} reported on cosmological constraints from the XMM Omega project which appeared to confirm a $L\propto T^3$ relation, however with a possible break in the LT slope at low redshifts. The need to modify the LT relation comes mainly from their temperature distribution function result $\Omega_m\approx 0.85$ which is in gross violation of the concordance value determined by many other measurements.

On the future AMI detector {\em Th. Culverhouse} reported work on models for estimating the confusion noise and thus to obtain an improved cluster selection function. {\em W. Grainger} reported on the AMI construction. An antenna array of small dishes for thermal SZ detection was now commissioning; an array of large dishes for point-source removal would also be ready to take data in 2005.

\subsection{Large Scale Simulations}
{\em P. Mazzotta} compared galaxy cluster temperatures from X-rays with emission-weighted temperatures from hydrodynamical N-body simulations of clusters, finding that they were in disagreement, particularly so in thermally highly inhomogeneous structures. The projected spectroscopic temperature is not a well defined quantity. An approximate agreement could be restored by introducing a phenomenological "spectroscopic-like temperature". {\em G. Murante} had simulated 200M hot and cold baryons in a cosmological tree + SPH in order to study the ICL (intra-cluster light), its dynamics, profile and cause. Evidence for ICL was found in cosmological hydrodynamical simulations, the more massive galaxies the greater amount of ICL. Stars in the cluster field seem to be older than those in galaxies. {\em M. Trenti} studied analytical models and N-body simulations of stellar systems under collisionless collapse, starting from uniformly and spherically distributed cold clumps, and resulting in elliptical galaxies. {\em L. Teodoro} presented N-body simulations on scales $10^{10}-10^{15}$ M$_{\odot}$ of $10^7-10^8$ DM particles with the purpose of obtaining convergence tests of the halo mass function, N-point correlation functions, the power spectrum and pairwise velocities.

\subsection{Neutrino Cosmology}
The only neutrino detector discussed was ANTARES, about which {\em V. Kudryavtsev} told that the construction had been started. The detector which is sensitive to energies $E_{\nu}>1$ TeV will be installed in 2006. 

{\em K. Abazajian} summarized neutrino mass limits and presented a simulation study of the influence of neutrinos on galaxy DM halos using analytic and numerical methods. {\em V. Lukash} studied cosmological parameters, deriving the cosmic degeneracy between $\Omega_{\Lambda}$ and the number of neutrino families, $f_{\nu}$. This degeneracy is inherent in the evolution of galaxy clusters and in other LSS data. His parameter values agreed with other published determinations. {\em E. Guendelman} proposed to explain DM by a new kind of neutrinos:  regular fermionic matter and a new dark fermionic matter would be different states of the same primordial fermion field. 

\section{Astronomy and Star Formation}
There is a conflict in cosmological distance measurements by standard rulers. {\em M. Kunz} showed that the luminosity distance $d_L(z)$ to SNe Ia, the angular diameter distance $d_A(z)$ to radio galaxies, and the  reciprocity relation $d_L(z)=(1+z)^2 d_A(z)$ do not agree. Could this be a sign of photon number violation or of $\gamma$-axion mixing ?

{\em T. Wyder} described the first results from the GALEX UV telescope on the properties of UV-luminous galaxies: the UV 2--point correlation function, UV extinction in galaxies, the star formation rate in the far and near UV and its evolution for $z<0.2$. The luminosity density in the local universe is observed to be lower than prevoiusly estimated. {\em S. Andreon} described the first results from the XMM-LSS survey and their future prospects. The
survey has found$\approx 75$ clusters with $0.3<z<1.05$. A passive population with early star formation only is seen in $2<z<5$, and an active population with star formation at all $z$. {\em M. Doherty} reported on near-infrared spectroscopy of $\approx 70$ galaxies in a recently started surey of star formation rates at $z=1$. {\em P. Rosati} working with the HST/ACS VLT reported that massive early clusters were in place already at $z=1.2$, the bulk of their stars having been formed at $z=2-3$. This does not exclude additional recent star formation. The conclusion is that hierarchical models are in trouble.

{\em C. Harrison} reporting on work with FLAMES and VLT UT2, noted that in early-type galaxies the relative ages and metallicities of stellar populations are degenerate. Developing an index more sensitive to ages, based on H$_{\beta}$ line strengths, and another index more sensitive to metallicity, based on Mg Fe line strengths, this degeneracy can be largely removed. {\em M. Viel} recovered cosmological parameters from the LUQAS sample of the Ly $\alpha$ forest. He noted that various systematic errors affect $dn_s/dk$ so that one could only conclude that there was no evidence for a tilt or for a non-vanishing $dn_s/dk$. In these data $\sigma_8$ is high. A robust estimator is the flux bispectrum.

{\em E. Pierpaoli} excluded that the reionization at $z=17$ claimed by WMAP could have been caused by star ionization. What kind of decaying particle could it then have been? Heavy neutrinos with the decay $\nu\rightarrow\pi e^+$ are not sufficient. CMB requires particles with long lifetime, limited abundance and limited decay rate at high $z$ in addition to the ionization limit. {\em R. Thompson} combined IR and optical HST observations in the ultra-deep field, $z$ approaching $\approx 11$, in a preliminary study of star formation rates and the LF. The intention is to put constraints on the objects that reionize the Universe.
 
\section{Cosmic Rays}
Selected experiments and results were reviewed by {\em M. Haas} and {\em M. Prouza}: the break in the spectrum slope at the "knee" and the "ankle", the GZK-cutoff understood to be caused by CMB absorption, the antiprotons consistent with being secondary. Ultra-high energies beyond the GZK limit will be studied by Auger. {\em M. Prouza} modelled the Galactic magnetic field including toroidal and poloidal turbulence field components. Simulating ultra-high-energy cosmic ray magnitudes and deflections and investigating the case that the cosmic rays are protons or Fe nuclei, he concluded that the incoming flux into the Galaxy could be anisotropical and that heavier nuclei appeared to come from the M87. 

\section{Cosmic Microwave Background}
The conference had hoped for the WMAP 2nd year results, but these were not yet released. {\em F. Hansen} reported an asymmetry in the power spectrum on large scales ($\ell=2-40$): the Northern hemisphere exhibits less fluctuations than the Southern hemisphere, a $3\sigma$ effect. It was not clear whether this reflected non-Gaussian signals, systematic errors, foreground or new physics. The multipoles $\ell=2-4$ are in any case foreground contaminated. The bispectrum of WMAP is consistent with Gaussianity, whereas {\em J. Medeiros} reported that the non-Gaussian signals seen in the COBE/DMR bispectrum were now understood: a combination of systematic effects and pixelization.

A technical description of ARCHEOPS, improved results and future plans to measure temperature anisotropies and dust polarization were presented by {\em M. Tristram} and {\em N. Ponthieu}. ARCHEOPS now covers a larger multipole range, $\ell=10-700$, and a larger sky coverage.  From an analysis based on the cross power spectrum method, a preliminary power spectrum was shown, in which the second acoustic peak was fairly high. Polarization has been detected in dense dust clouds ($\approx 10\%$) and in diffuse regions near the Galactic plane ($3-5\%$). The polarization orientation is compatible with a Galactic magnetic field aligned with the spiral arms.

The latest results and future plans for VSA were discussed by {\em K. Grainge} and {\em R. Battye}. The power spectrum is measured and cosmological parameters are estimated from data at multipoles $\ell=150-1600$, while future upgrades planned to reach $\ell> 2000$. VSA addresses the physics of recombination, fluctuation damping, the functional form of the scalar spectral index $n_s(k)$, other cosmological parameters, and the presence of cosmological defects. The WMAP(first year)+VSA+CBI+ACBAR data prefer, at $2\sigma$, a running spectral index. Inclusion of LSS data from 2dFGRS weakens this conclusion.

{\em B. Rusholme} described the South Pole-based bolometric polarimeter QUaD (QUEST And DASI) which whill become operational this year. Its task is to map $E$- and lensed $B$-mode polarization on scales $>4'$, focusing on low foreground areas of the Southern sky. {\em C. Rosset} noted that CMB polarization is rich of cosmological information. Important limitations in PLANCK are the foregrounds and systematic effects of different instrumental parameters on the $E$- and $B$-mode polarization power spectra, for instance the $E\rightarrow B$ leakage effect. {\em A. Taylor} discussed the design and expected performance of the two new complementary instruments CLOVER and BRAIN for measuring the $B$-mode polarization. CLOVER is a focal plane array which will measure multipoles $\ell=20-1000$. BRAIN is the first bolometric interferometer.

Progress in the design of software for CMB analysis was presented by several speakers. {\em Ch. M\"{u}ller} extended the CMBEasy software with a Markov chain Monte Carlo simulation and data analysis package. {\em A. Slosar} described the development of a fast likelihood  routine for evaluating low WMAP quadrupoles and octupoles by direct inversion of the theoretical covariance matrix. This projects out foreground Galactic contaminants, and results in a increase of power at low multipoles, favoring lower values of $\Omega_m$. {\em J. McEwen} presented fast algorithms for continuous-directional spherical wavelet transform, permitting the analysis of non-isotropic signals. Some possible deviation from Gaussianity has been detected in WMAP data.

\section{Primordial Universe}

\subsection{Inflation}
{\em W. Kinney} reviewed inflation zoology and the implications of WMAP on inflation models. A MC generation of 1M inflation models in the $n_s,\ r$-space ($r=$tensor/scalar) showed WMAP to be consistent with many models, not only "vanilla" ($r=0,\ n_s=1$), with $n_s$ running or not, but there is no evidence for tensors. Potentials of the form $\lambda\phi^4$ are under pressure but not quite ruled out yet. One concludes that the primordial fluctuations are indeed adiabatic and acausal. Hybrid inflation is favored, but it does not challenge the paradigm of single-field inflation which satisfies all CMB and LSS constraints. {\em J. Garcia-Bellido} agreed with this in his review of isocurvature perturbations which are entropy perturbations and entail changes in particle numbers.  Their contribution does not significantly improve fits with a pure adiabatic model. {\em J. Dunkley} used CMB and LSS data to measure the amplitude of correlated non-adiabatic fluctuations. She found a broader range of possible structure formation initial conditions, and concluded that a high isocurvature fraction could perhaps be allowed in a different inflationary model. The present high precision values obtained for cosmological parameters are only valid in the adiabatic model, whereas isocurvature model tests are needed. 

 {\em K. Dimopoulos} described a hybrid inflationary potential with two modular scalar fields $\Phi,\ \phi$ which characterize SUSY flat directions. The flatness is lifted by SUGRA corrections. The potential contains terms of type $\Phi^2,\ \Phi^2\phi^2,\ \phi^2,\ \phi^4$ and has an unstable saddle point. The first stage of inflation is fast-roll, depending on curvature. The second stage is locked inflation on the saddle point, followed by a third stage of fast-roll which solves the horizon and flatness problems. The observed superhorizon spectrum of curvature perturbations is due to a curvaton field $\sigma$ with no influence on inflationary dynamics. The $\sigma$ ameliorates the tuning problems of inflation model-building and allows inflation to be achieved without the use of flat directions. After inflation the $\sigma$ decays into the hot Big Bang thermal bath.

{\em M. Weinstein} described a simple formalism for setting up a fully quantum calculation of the CMB fluctuations, exactly solvable in de Sitter space. In the infinite past and future the solution is classical, in the finite past one does not obtain a purely exponential expansion. One might be able to observe this, or to bound the earliest time at which one is free to set initial conditions on the inflation. This theory predicts possible corrections to $\delta\rho/\rho$ and a back-reaction for the scale factor. 

{\em S. Winitzki} discussed the fractal structure of inflationary spacetime, noting that models of eternal inflation (e.g. bubble nucleation) allow the existence of a fractal set of points that never thermalize. This has consequences for hybrid inflation and for holography. According to {\em J.C. Hwang}, the second-order relativistic Zel'dovich approximation to perturbations in a cosmological post-Newtonian formulation are closer to the linear approximation than to the non-linear case.
 
\subsection{Cosmic Strings}
Cosmic string formation in SUSY GUTs do not appear to contribute to CMB anisotropies, but {\em M. Landriau} claimed that they could nonetheless be present at a lower energy scale. In his preliminary calculation their effect was further reduced to a level $<6\%$ with a large uncertainty. {\em M. Sakellariadou} explored the full parameter space of SUSY GUTs and concluded that at most $9\%$ could be tolerated, their effect being maximal at low multipoles where systematic errors also are maximal and where tensors might also confuse the issue. One is therefore approaching the point where the string contribution is arbitrarily close to $0\%$. Strings are like dinosaurs: one cannot prove that they are extinct, but one can set upper limits to the probability of finding one.

In view of this, the talk of {\em J. Rocher} is highly interesting. He examined all possible spontaneous breaking patterns from SUSY GUT Lie groups down to the standard model, concluding that all patterns lead unavoidably to cosmic strings. Thus one has an important no-go theorem: a demonstrated non-existence of cosmic strings rules out any simple relation between inflation and SUSY GUT-symmetry  breaking as the origin of Standard Model particle physics.  Thus one has to turn to more complicated models where strings do not appear, as for instance hybrid inflation.

\subsection{Brane Cosmology}
Given the lack of a quantum gravity theory, M-theory which needs extra dimensions and branes holds some promises. {\em R. Maartens} reviewed braneworld phenomenology which has many new features which future precision cosmology can constrain: Ka\l uza--Klein models, moduli fields, holography, shadow matter. Ordinary matter is constrained to the D-brane, gravity acts in the full-dimension bulk space but is localized close to the D-brane. Thus one has a different handle on DM, which is not a particle or sparticle with standard interactions, but which exists on a different brane and is felt only by its gravitational interactions. Simple RS (Randall--Sundrum) models are OK so far, but next one has to start to calculate CMB power spectra for the various alternatives. 

{\em R. Holman} considered brane preheating by the radion, a massless 4-dimensional field which encodes the physical separation between branes in compact RS models. It does not mix with gravitons, but it couples directly to particles on both branes. It can be considered a geometrical degree of freedom, having interesting applications to baryogenesis and perhaps the generation of gravity waves. {\em A. Mennim} studied various approaches to calculate the evolution of cosmological perturbations in the single-brane RS model.

In another gravity model {\em S. Davis} described the implications of the Gauss--Bonnet curvature term for brane world 5-dimensional gravity. Such terms may allow singularity problems to be avoided, but may also lead to new instabilities, ghosts and tachyons. According to {\em A. Lopez Maroto}, massive brane fluctuations, branons, are natural DM candidates. They are stable, have weak interactions, and can be searched for in colliders and other particle experiments. A critical question is why they haven't been discovered yet.

\subsection{Fine-structure Constant}
{\em C. Martins} reviewed the claims for and against a time-varying fine-structure constant $\alpha$. While the claims for time-variation prevailed, several attempts were made to link the effect to DE, notably to a time-varying $w_{\phi}$. It was obvious, however, that any fit of a time-function to the value $\Delta\alpha/\alpha\approx 10^{-5}$ from QSO absorption lines at lookback times $0.2-0.9$, and to the Oklo natural reactor value $\Delta\alpha/\alpha\approx 10^{-7} -10^{-8}$ at lookback time 0.14 would be highly contrived. It then came as a relief that new data from the WMAP and VLT now contradict the QSO data.  The systematic errors must be huge. Thus all theoretical modelling is premature. 

\section*{Acknowledgments}
My compliments to Tran Thanh Van and his collaborators for having organized this successful meeting. I feel very honored for having been given the task to summarize it.

\end{document}